\documentstyle[11pt,paspconf,epsf]{article}

\newcommand{\cd}{{\it cd }}
 
\newcommand{\Msun}{\, {\rm M}_{\odot}}
\newcommand{\Zsun}{$\, Z_{\odot}$\,}
\newcommand{\AuA}[2]{{\it A{\rm \&}A\/}, {\bf #1}, #2}
\newcommand{\APJ}[2]{{\it ApJ\/}, {\bf #1}, #2}
\newcommand{\APJSS}[2]{{\it ApJS\/}, {\bf #1}, #2}

\newcommand{\AJ}[2]{{\it AJ\/}, {\bf #1}, #2}

\newcommand{\PASJ}[2]{{\it PASJ\/}, {\bf #1}, #2}

\newcommand{\MN}[2]
   {{\it MNRAS\/}, {\bf #1}, #2}

\begin{document}
\title{Chemical Evolution of Galaxies \\
and the Relevance of Gas Processes}
\author{Gerhard Hensler, Andreas Rieschick}
\affil{Institut f\"ur Theoretische Physik und Astrophysik, 
Universit\"at Kiel, \\
Olshausenstr.\ 40, D--24098 Kiel, Germany, \\
email: hensler@astrophysik.uni-kiel.de} 

\begin{abstract}
Since stellar populations enhance particular element abundances
according to the yields and lifetimes of the stellar progenitors,
the chemical evolution of galaxies serves as one of the key tools
that allows the tracing of galaxy evolution.
In order to deduce the evolution of separate galactic regions 
one has to account for the dynamics of the interstellar
medium, because distant regions can interact by means of large-scale 
dynamics. To be able to interpret the distributions and ratios 
of the characteristic elements and their relation to e.g.\
the galactic gas content, an understanding of the dynamical effects
combined with small-scale transitions between the gas phases 
by evaporation and condensation is essential.
In this paper, we address various complex signatures of chemical
evolution and present in particular two problems of abundance 
distributions in different types of galaxies: the discrepancies 
of metallicity distributions and effective yields in the different 
regions of our Milky Way and the N/O abundance ratio in dwarf galaxies.
These can be solved properly, if the chemodynamical prescription 
is applied to simulations of galaxy evolution. 
 
\end{abstract}


\section{Introduction}

For the stellar populations of our Milky Way Galaxy
(MWG) - the halo, the bulge, and the disk (thick plus
thin disk) - the fundamental questions that have to be addressed
are: When, how and on what timescales did the 
Galactic components form, and was there any connection between them?
If yes, simultaneously or sequentially? One possible approach to 
disentangle the evolutionary scenario is to look for evolutionary 
signatures in age, dynamics, and chemistry of long-lived 
stars, in the stellar populations within our
MWG. At present, two major and
basically different strategies for modelling galaxy evolution can be followed:
dynamical investigations which include hydrodynamical simulations of
isolated galaxy evolution and of protogalactic interactions reaching from
cosmological perturbation scales to direct mergers, and, on the other hand,  
studies which neglect any dynamical effects but consider either the whole
galaxy or particular regions and describe
the temporal evolution of mass fractions and element abundances
in detail. 

For the case of a closed box a linear relation 
between the time-dependent metallicity $Z(t)$ and the 
initial-to-temporal gas ratio $ln[M_{g,0}/M_g(t)]$ follows
analytically, where the slope is determined by
the yield $y$, i.e.\ the metallicity release per stellar population. 
Deviations from this simple relation are explained by lower ``effective''
yields $y_{\sf eff}$ due to outflow of metal-rich gas from the 
(now open) volume or infall of low-metallicity (presumably 
primordial) gas. Such dynamical effects can only be properly treated 
if simulations can account for the energetics, the composition, 
and the dynamical state of the galactic gas, as well as the relevant 
interchange processes, in a self-consistent manner.
This includes the pollution of the different gas phases with
characteristic elements by means of various stellar mass-loss
mechanisms, the gas phase transitions, and the 
self-consistent large-scale streaming motions of hot intercloud gas 
(ICM) or cool gas (CM) that falls in from a reservoir of 
protogalactic gas. It follows
that galactic regions and components experience mutual 
dynamical interactions and their evolutions are not decoupled.  

Let us consider a number of chemical peculiarities of different 
kinds of galaxies which justify the use
of sophisticated multi-phase dynamical descriptions of the 
interstellar medium (ISM) in studies of the global scenario of galaxy 
evolution:

\noindent
1) In the solar vicinity at least three severe problems arise by 
considering the metallicity distribution of F or G dwarfs, 
stars that live sufficiently long to trace the evolution 
of a galaxy: The age-metallicity relation, $y$ of the 
metallicity distribution and, finally, the lack of metal-poor G dwarfs 
(the well-known G-dwarf problem). Various influences on the evolution 
of the solar neighbourhood have therefore been invoked by different 
authors and are applied by artificial parametrizations ranging from 
time-dependent accretion of pristine halo gas to temporal variations 
of the stellar initial mass function (IMF). Although they lack
self-consistency, the results can often provide very helpful basic 
insights into influences of distinct effects. 

\noindent 
2) It is also important to explain the observed differences of 
$y$ in bulge, disk and halo of the MWG (Pagel 1987).

\noindent
3) There is now much evidence that metal-absorption line systems (ALS) in
QSO spectra arise from the gaseous halos of forming galaxies. In spite of
the lack of knowledge of their dynamical properties and their origin,
as a working hypothesis it is assumed
that the early hot halo gas is produced by supernovae typeII 
(SNeII), because the observed abundance ratios agree well with 
SNII yields (Reimers et al.\ 1992). The metallicity and radial extent
of ALS should therefore appear automatically in galaxy models as the 
result of chemical evolution.

\noindent
4) Dwarf galaxies (DGs) present a variety of morphological types.
Their structural and chemical properties differ from those of giant 
galaxies. The dwarf irregular galaxies (dIrrs) appear 
with lower $Z$ at the same gas fraction as gSs. This implies 
that the metal-enriched gas from SNeII was lost from the galaxy rather 
than astrated (Larson 1974, Dekel \& Silk 1986).
Yet many dwarf spheroidals (dSphs) which represent the low-mass end
of DGs show not only a significant intermediate-age 
stellar population, but also more recent SF events (Smecker-Hane et al.\ 1994,
Han et al.\ 1997) with increasing metallicity, 
indicating that gas was kept in the system.

\noindent
5) gSs and DGs differ significantly in their N/O ratios. 
A fundamental explanation is needed to explain why gSs like the MWG reach 
higher ratios at larger O abundances than DGs, while N/O is almost restricted to
around -1.5 for DGs over a wide range in O (see fig.3).

\section{The Chemodynamical Treatment}

For systems and sites of low potential energy we know from empirical 
studies and theoretical investigations that the ISM is on average 
held in balance by counteracting processes like heating and cooling, turbulence 
and dissipation (Burkert \& Hensler 1989, Hensler et al.\ 1998b). 
Since these processes are 
non-linearly coupled, the effect of neglecting one of them will 
alter the evolution completely. A number of studies of
self-regulated SF exist with particular attention to the influences
of stellar radiation, supernova explosions, and the 
evaporation/condensation balance between the two chemically
and dynamically distinct gas phases, the cloudy medium (CM)
and the hot intercloud medium (ICM) 
(Franco \& Cox 1983, Ikeuchi et al.\ 1984, McKee 1989, 
Bertoldi \& McKee 1995, K\"oppen et al.\ 1995,1998). 
The evolution of DGs is self-regulated and
determined by large-scale outflows (Hensler et al.\ 1993,1998a). 
External effects like extended DM halos, the IGM pressure, etc.
(Vilchez 1995) could cause the morphological differences 
of DGs by regulating the otherwise unbound hot gas that can be held in the 
galactic halo. The gas could then either be stripped off, or it
cools and recollapses, igniting subsequent SF.

Self-regulation with SNII energy deposition can also characterise
the structure and evolution of
galactic disks which reach a lower effective gravitational potential
in rotational equilibrium
(Firmani \& Tutukov 1992; Burkert et al.\ 1992, Rosen \& Bregman 1995).

To approach global models of galaxy evolution
which yield the structural differences and details, adequate
treatment of the dynamics of stellar 
and gaseous components is essential. At least the 
following processes have 
to be taken into account: SN, SF, heating, cooling, stellar mass loss, 
condensation and evaporation. This includes the treatment of
the multi-phase character of the ISM as well as  the star-gas 
interactions and phase transitions. Since gas and stars evolve 
dynamically, and because several processes both depend on their metallicities
and also influence the element abundances in each component,
these models are called {\bf chemodynamical} (\cd). 

It must be emphasized, however, that the number of free parameters 
in the \cd scheme is not 
large but actually smaller than in multi-zone models. The allowed ranges of 
parameter values
are strongly constrained, either because they are theoretically 
evaluated (like e.g.\ evaporation and condensation), empirically 
determined (like e.g.\ stellar winds), or because they force 
self-regulation in a way that is independent of the parameterization. 
Because of limited space we refer the interested reader to 
more comprehensive descriptions of the \cd treatment and to different 
applications (non-rotating galaxies: Theis, Burkert \& Hensler 1992, 
Hensler, Burkert \& Theis 1993, Hensler, Gallagher \& Theis 1998a; 
vertical settling of the galactic disk: Burkert, Truran \& Hensler 1992;
disk galaxies: Samland \& Hensler 1996; 
the MWG: Samland, Hensler \& Theis 1997 (SHT97), Samland 1998; 
dwarf galaxies: Hensler \& Rieschick 1998, also section 5).

\section{The Milky Way's Chemical Evolution}

As a striking success of the \cd treatment we will first briefly present
some of the results from a published model (SHT97) which can be
compared to the observational features of
the MWG. The model starts from an isolated spheroidal, rotating
but purely gaseous cloud with a mass of
$3.7 \cdot 10^{11} \Msun$, a radius of 50 kpc, and an angular momentum
of about $2 \cdot 10^7 \Msun$ pc$^2$ Myrs$^{-1}$, 
corresponding to a spin parameter $\lambda$=0.05. We assume that the 
protogalaxy consists initially of CM and ICM with a density distribution of 
Plummer-Kuzmin-type (Satoh 1980) with 10 kpc scalelength.
The initial CM/ICM mass division (99\%/1\%) does not 
affect the later collapse, because the onset of SF determines the physical 
state within less than 10$^7$ years.

\begin{figure}
\plotfiddle{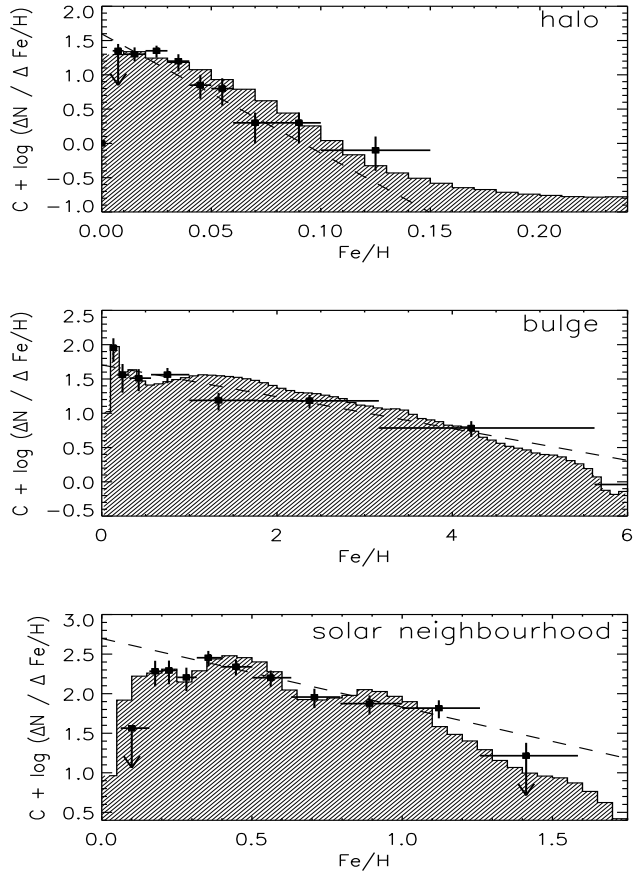}{110truemm}{0}{50}{50}{-200}{-50}
\caption{Differential iron distribution of stars in the halo, K giants in the 
         bulge, and G dwarfs in the solar neighbourhood in comparison with  
         observations. The dashed lines show the results of simple one zone 
         models with effective yields of 0.025 \Zsun (halo), 1.9 \Zsun 
         (bulge) and 0.5 \Zsun (disk)(from SHT97).}
\end{figure}

While the evolutionary phases are described in detail in SHT97, here we
wish to emphasize the striking agreement of the \cd model 
after 15 Gyrs with first the metallicity distributions 
of the halo, the bulge, and the 
solar vicinity (fig.1), and secondly the radial oxygen gradient within 
the disk (fig.2). Convincingly one single \cd model
reproduces the different (otherwise implausible) $y_{\sf eff}$ in the
different regions and demonstrates that they result simply from
large-scale streaming effects of the hot gas. 
The ICM is produced by SNeII in overpressure to its surrounding ISM 
and expands until its cooling leads to condensation, i.e.\ the
phase transition to the CM.
In addition, the model shows agreement with the MWG structure,
i.e. gas-star content, baryonic mass distributions, velocity distributions,
PN and SN rates. From this agreement it may be safe to
assume that the temporal evolution of the model is reliable, e.g.\ 
the formation epochs of the components and the temporal 
variation of the radial metallicity gradient (see SHT97).

\begin{figure}
\plotfiddle{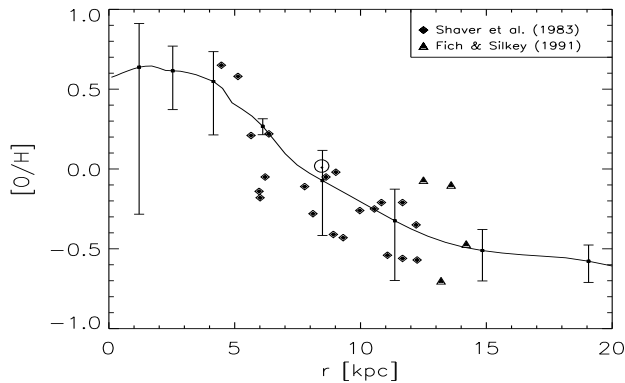}{60truemm}{0}{50}{40}{-150}{0}
\caption{Oxygen gradient of the CM in the equatorial plane 
         of a chemodynamical Milky Way model after $15\cdot10^9$ 
         years. The error bars indicate the local fluctuations in the model. 
         Observational data of the sun and of H{\sc II} regions 
         (rhombi, triangles) are plotted for comparison (from SHT97).}
\end{figure}

\section{Small and Large-scale Mixing Effects between the Interstellar 
Gas Phases and what can Abundance Ratios tell us about?}

The enormous energy release by massive stars from their combined
wind, radiation and SNII explosion leads to violently
expanding hot gas bubbles. They act dynamically
on the ambient ISM by sweeping it up
and squeezing it into dense shells, which break up due to dynamical
instabilities. Since massive stars have peeled off their unprocessed
shell material during the H-main sequence lifetime, they expel their 
nucleosynthesized products before and during their 
Wolf-Rayet phase and even more intensely by SNII explosion. Due to its
rapid expansion the hot gas engulfs the denser cool clumps and clouds. 
The effect is twofold: 
First, as it passes the clouds the ICM significantly perturbs their shape 
and surface (Elmegreen, this volume). Secondly, the contact interface
between CM and ICM is blown up by heat conduction. If the cloud is 
able to get rid of the diffused energy, hot gas can condense onto 
its surface; if not, the cloudy material evaporates from the surface 
and immigrates into the ICM. Reasonably, this mass exchange by means of 
evaporation/condensation is a self-regulated process in static media
(K\"oppen et al.\ 1998). Streaming motions, however, can alter this 
picture and can lead to runaway behaviour in either direction.
Certainly this mechanism causes a highly efficient small-scale 
mixing between the gas phases and by this homogenizes abundances
on the local scales of massive star associations.

Indeed, Kobulnicki (this volume) reports the non-detection
of any sizable O, N, and He anomalies from H{\sc ii} regions in the 
vicinity of young super starclusters in starburst DGs with one exception,
NGC 5253, which reveals a central N overabundance.

\begin{figure}
\plotfiddle{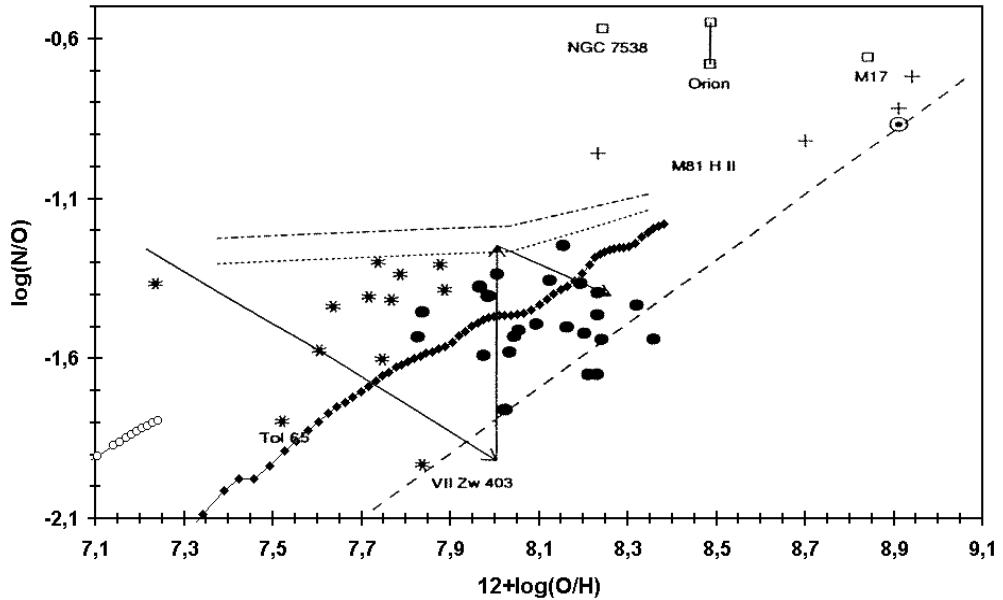}{85truemm}{-90}{50}{45}{-220}{260}
\caption{Evolutionary tracks of 2d chemodynamical models for 10$^9 \Msun$
         galaxies with (long full dotted curve) and without DM halos 
         (open circles) in comparison with 
         N/O vs.\ O/H measurements of irregular galaxies and two
         chemical evolutionary models by Matteucci \& Tosi (1985) 
         (upper lines) and a simple model (arrows; see Garnett, 1990). }
\end{figure}

While C and N are mainly contributed to the CM by PNe from
intermediate-mass stars, O and Fe are the dominant tracers of SNII
and SNIa ejecta, respectively. Since the mixture of e.g.\ N and O
can only result from phase transitions between CM and ICM, the N/O
ratio allows to make qualitative deductions about the mixing direction 
and its temporal efficiency. Additionally, 
its radial distribution provides an insight into dynamical effects of 
the ISM. As mentioned in the introduction (point 5) the N/O ratio is smaller
in DGs than in gSs by almost 0.7 dex, while O is lower by
one order of magnitude. The averaged N/O value for DGs (see fig.3) 
at -1.46 (Garnett 1990) lies only 0.2 dex below the ratio determined from 
metal-dependent yields (Woosley \& Weaver 1995) for \Zsun
and integrated over a Salpeter IMF. The almost linear dependence 
of the nitrogen production on $Z$ allows for even smaller N/O
in DGs because of the generally lower $Z$, which explains the
observed scatter to even smaller ratios. 

In order to study and compare abundances and structural signatures in \cd 
models of DGs, we have performed 
2d simulations of rotating gaseous protogalactic clouds
for a large range of initial masses with and
without dark matter halos. The density distribution is again of
the Plummer-Kuzmin-type with the same spin parameter as the 
above-mentioned MWG model. Here we discuss 10$^9 \Msun$ DG models
starting with a 2 kpc scalelength.
Fig.3 presents a diagram for the N/O vs.\ O/H ratios of DGs and gSs 
(also the solar value) compared with evolutionary tracks by 
different authors. In contrast to the other models shown (see
Garnett 1990) which begin and partly remain at too large 
N/O ranges, both our \cd models commence at very low values due to 
the delayed 
nitrogen release by PNe and the lower N production in massive stars at
low metallicities. The most evolved track (10$^9 \Msun$ with DM)
rises rapidly and reaches N/O of -1.8 after 1 Gyr and -1.6 after 2 Gyrs, 
respectively. This small value results numerically
as the yield ratio, but as N and O enrich different gas 
phases, an almost ideal mixing of CM and ICM is required. 
Since the hot material cannot fully condense onto the
clouds, an almost total evaporation of the CM in the vicinity of the 
SF and SNII explosion sites must be invoked. Reasonably, the N mixes 
perfectly in this case with O in the ICM to an almost constant 
abundance ratio
and expands over larger distances within the DG. Due to its cooling
the ICM forms new condensations of CM where the abundances 
are observed in H{\sc ii} regions at a constant value 
(Kobulnicki, this volume). In the case of phase transition by means 
of condensation, only parts of the ICM
(and therefore of the O content) is incorporated into the CM, which
leads to higher N/O ratios, but also reveals inhomogeneous N/O
distributions. Larger gravitational potentials and
resulting mass densities in more massive galaxies lead to a faster
cooling of the ICM and significantly higher condensation rates which
produces a larger N/O. With the same differential mixing processes the 
observed C/O tendency of DGs (Garnett et al.\ 1995) can also be explained
(Rieschick \& Hensler, in preparation).

\section{Conclusion}

Because of limited space here, we could only briefly demonstrate that various 
structural and chemical agreement with observations can be achieved 
self-consistently by global evolutionary \cd models. If the \cd 
prescription is applied, 
important physical processes like large-scale coupling of different 
galactic regions by dynamical interactions as well as small-scale
mixing effects between the gas phases are adequately taken into account,
and this substantially fixes the element abundances. Abundances
can serve as reliable diagnostic tools of galaxy evolution and
provide a chance to deconvolve it in detail, if studies couple the
above-mentioned gas processes with the dynamics of gas and stars as
well as with their mutual interactions.

\acknowledgments 
G.H. is very grateful to the organizers of the conference for their 
invitation and their kindest hospitality. The authors
achnowledge gratefully cooperations and discussions with J.K\"oppen, 
M.Samland and Ch.Theis and their contributions to the field of chemodynamics. 
We also thank the referee Mike Edmunds for the careful reading of the 
text and his comments for clarification. 
This work is supported by the {\it Deutsche Forschungsgemeinschaft} (DFG) 
under grant no.\ He 1487/5-3 (A.R.), 
the participation at this meeting by one of us (G.H.) under grant 
no.\ He 1487/21-1. The numerical 
calculations are partly performed at the computer centers RZ Kiel, 
ZIB Berlin, and HLRZ J\"ulich. \\ [-0.8cm]


\end{document}